\documentstyle[12pt]{article}
\begin{document}

\vskip 1.0cm \centerline{\LARGE\bf Nonexistence of
time-reversibility in} \vskip 2pt \centerline{\LARGE\bf
statistical physics} \vskip 10pt \centerline{C. Y. Chen}
\centerline{Department of Physics, Beijing University of
Aeronautics} \centerline{ and Astronautics, Beijing, 100083,
P.R.China}

\vskip 10pt
\begin{center}
\begin{minipage}{12cm}
{\noindent {\bf Abstract:} Contrary to the customary thought
prevailing for long, the time reversibility associated with
beam-to-beam collisions does not really exist. Related facts and
consequences are presented. The discussion, though involving
simple mathematics and physics only, is well-related to the
foundation of statistical theory. }
\end{minipage}
\end{center}


\section{Introduction}

More than one hundred years ago, the debate concerning time
reversibility arose in a confusing way: Boltzmann derived his
kinetic equation from the time reversibility of mechanics while
the equation itself was of time irreversibility\cite{reif}. Even
today, though a long time has passed and countless papers in the
literature have revealed a variety of aspects of the issue,
paradoxical things still bother some of us\cite{chen02}.

Here, it will be shown that the real problem of Boltzmann's theory
is related not to the time irreversibility assumed by it, but to
the time reversibility assumed by it. To make the topic more
intriguing and more profound, the investigation will manifest that
any attempt to formulate the `true distribution function' will
fail in the ultimate sense.

\section{Particle-to-particle and beam-to-beam collisions}
Before entering the detailed discussion, it is essential to
establish distinction between particle-to-particle collisions and
beam-to-beam collisions.

A particle-to-particle collision involves two individual
particles. The time reversal symmetry of it has been fully
elucidated in classical mechanics and it says that if the
collision expressed by $({\bf v}_1,{\bf v}_2) \rightarrow ({\bf
v}'_1, {\bf v}'_2)$ is physically possible, where ${\bf v}_1$,
${\bf v}_2$ are respectively the velocities of the two particles
before the collision and ${\bf v}'_1$, ${\bf v}'_2$ after the
collision, then the inverse collision expressed by $(-{\bf v}'_1,
-{\bf v}'_2)  \rightarrow (-{\bf v}_1,-{\bf v}_2)$ is also
physically possible. This kind of time reversibility is not truly
relevant to the subject herein and we shall not discuss it too
much in this paper.

Beam-to-beam collisions, involving particle densities or
distribution functions, are of great significance to statistical
mechanics. For instance, Boltzmann's theory treats $f({\bf
v})d{\bf v}$ as a beam and derives its formulation on the premise
that certain types of time-reversibility are there.

However, we happen to realize that no time-reversibility of any
form can be defined in the context of Boltzmann's theory. This
conclusion is surprising, seems very imprudent and directly
contradicts what has been embedded in our mind. In view of such
strong resistance, it is felt that a very clear and very detailed
discussion should be given. In this section the subject will be
studied intuitively and in the next two sections mathematical
investigations will be presented.

\hspace{1.4cm} \setlength{\unitlength}{0.020in}
\begin{picture}(100,76)
\multiput(24,23)(-1,2){2}{\vector(2,1){22}}
\multiput(24,52)(-1,-2){2}{\vector(2,-1){22}}
\put(52,41){\vector(1,1){18}} \put(53,40){\vector(3,2){20}}
\put(51,42){\vector(2,3){14}} \put(52,34){\vector(1,-1){18}}
\put(53,35){\vector(3,-2){20}} \put(51,33){\vector(2,-3){14}}
\put(50,2){\makebox(0,8)[c]{\bf (a)}}

\hspace{-20pt}

\multiput(145,39.6)(1,2){2}{\vector(-2,1){22}}
\multiput(145,36.4)(1,-2){2}{\vector(-2,-1){22}}
\put(171,59.5){\vector(-1,-1){18}}
\put(174,53.5){\vector(-3,-2){20}}
\put(166,63.5){\vector(-2,-3){14}} \put(171,16){\vector(-1,1){18}}
\put(174,22){\vector(-3,2){20}} \put(166,12){\vector(-2,3){14}}
\put(150,2){\makebox(0,8)[c]{\bf (b)}}
\end{picture}

\vskip0.1cm
\begin{center}
\begin{minipage}{12cm}
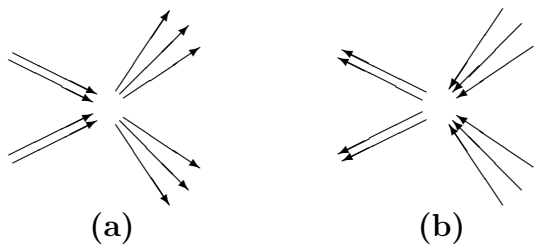
{Figure~1: A candidate for time reversibility of beam-to-beam
collision: (a) the original collisions; and (b) the inverse
collisions. }
\end{minipage}
\end{center}
\vskip0.2cm

Take a look at Fig.~1. Fig.~1a shows a process in which two beams
with two definite velocities collide and the produced particles
diverge in space. Fig.~1b illustrates the inverse process, in
which converging beams collide and the produced particles form two
definite beams. In no need of detailed discussion, we surely know
that the first process makes sense in statistical mechanics while
the second one does not.

\section{No time-reversibility in terms of cross sections}
In the textbook treatment, the time reversibility concerning
beam-to-beam collisions is expressed in terms of cross
sections\cite{reif}:
\begin{equation} \label{equality} \sigma({\bf v}_1,{\bf v}_2
\rightarrow {\bf v}_1^\prime, {\bf v}_2^\prime) = \sigma ({\bf
v}_1^\prime, {\bf v}_2^\prime \rightarrow {\bf v}_1, {\bf v}_2)
,\end{equation} where $\sigma({\bf v}_1,{\bf v}_2 \rightarrow {\bf
v}_1^\prime, {\bf v}_2^\prime )$ is defined in such a way that
\begin{equation}  \label{sigma1}
N= \sigma({\bf v}_1,{\bf v}_2 \rightarrow {\bf v}_1^\prime, {\bf
v}_2^\prime ) d{\bf v}_1^\prime d{\bf v}_2^\prime
\end{equation}
represents the number of type-1 particles emerging, after
collisions, between ${\bf v}_1^\prime$ and ${\bf v}_1^\prime+d
{\bf v}_1^\prime$ per unit incident flux and unit time, while the
type-2 particle emerges between ${\bf v}_2^\prime$ and ${\bf
v}_2^\prime+d {\bf v}_2^\prime$.

\hspace{6.9cm} \setlength{\unitlength}{0.014in}
\begin{picture}(200,143)

\put(-120,85){\vector(3,-1){84.5}}
\put(-120,85){\vector(1,1){27.5}}
\put(-120,85){\vector(1,0){56.5}}
\multiput(-91.5,113.5)(1.5,-1.5){36}{\circle*{1.2}}
\put(-58.2,85.3){\vector(1,-1){28.2}}
\put(-58.2,85.3){\circle*{3}}
\put(-106,111){\makebox(35,8)[l]{${\bf v}_1$}}
\put(-58,53){\makebox(35,8)[l]{${\bf v}_2$}}
\put(-90,77){\makebox(35,8)[c]{${\bf c}$}}
\put(-55,68){\makebox(35,8)[c]{${\bf u}$}}

\put(-80,28){\makebox(0,8)[c]{\bf (a)}}

\hspace{-2.2cm} \put(98,28){\makebox(0,8)[c]{\bf (b)}}
\put(46,85){\vector(1,0){55.5}}
\multiput(102.0,85)(-1.0,-2.0){19}{\circle*{1.2}}
\multiput(102.0,85)(1.0,2.0){19}{\circle*{1.2}}
\put(108,85){\circle*{3}} \put(108,85){\vector(-1,-2){18.0}}

\put(139,102){\makebox(35,8)[l]{$({\bf v}_1^\prime)$}}
\put(45.5,85){\line(5,1){92}} \put(136.5,103.4){\vector(4,1){1}}


\put(46.5,85){\vector(2,1){72.5}}
\put(121,121){\makebox(35,8)[l]{${\bf v}'_1$}}


\put(46,85){\vector(1,1){28.5}}
\put(57,114){\makebox(35,8)[l]{$({\bf v}_1^\prime)$}}

\put(46.5,85){\vector(1,-1){36}}


\put(66,47){\makebox(35,8)[l]{${\bf v}_2^\prime$}}
\put(69,76){\makebox(35,8)[c]{${\bf c}'$}}
\put(88,64){\makebox(35,8)[c]{${\bf u}'$}}

\put(133,50){\makebox(35,8)[l]{$S$}} \put(143.00, 85.00){\circle
*{1.2}} \put(142.78, 89.18){\circle *{1.2}} \put(142.13,
93.32){\circle *{1.2}} \put(141.04, 97.36){\circle *{1.2}}
\put(139.54, 101.27){\circle *{1.2}} \put(137.64, 105.00){\circle
*{1.2}} \put(135.36, 108.51){\circle *{1.2}} \put(132.73,
111.77){\circle *{1.2}} \put(129.77, 114.73){\circle *{1.2}}
\put(126.51, 117.36){\circle *{1.2}} \put(123.00, 119.64){\circle
*{1.2}} \put(119.27, 121.54){\circle *{1.2}} \put(115.36,
123.04){\circle *{1.2}} \put(111.32, 124.13){\circle *{1.2}}
\put(107.18, 124.78){\circle *{1.2}} \put(103.00, 125.00){\circle
*{1.2}} \put(98.82, 124.78){\circle *{1.2}} \put(94.68,
124.13){\circle *{1.2}} \put(90.64, 123.04){\circle *{1.2}}
\put(86.73, 121.54){\circle *{1.2}} \put(83.00, 119.64){\circle
*{1.2}} \put(79.49, 117.36){\circle *{1.2}} \put(76.23,
114.73){\circle *{1.2}} \put(73.27, 111.77){\circle *{1.2}}
\put(70.64, 108.51){\circle *{1.2}} \put(68.36, 105.00){\circle
*{1.2}} \put(66.46, 101.27){\circle *{1.2}} \put(64.96,
97.36){\circle *{1.2}} \put(63.87, 93.32){\circle *{1.2}}
\put(63.22, 89.18){\circle *{1.2}} \put(63.00, 85.00){\circle
*{1.2}} \put(63.22, 80.82){\circle *{1.2}} \put(63.87,
76.68){\circle *{1.2}} \put(64.96, 72.64){\circle *{1.2}}
\put(66.46, 68.73){\circle *{1.2}} \put(68.36, 65.00){\circle
*{1.2}} \put(70.64, 61.49){\circle *{1.2}} \put(73.27,
58.23){\circle *{1.2}} \put(76.23, 55.27){\circle *{1.2}}
\put(79.49, 52.64){\circle *{1.2}} \put(83.00, 50.36){\circle
*{1.2}} \put(86.73, 48.46){\circle *{1.2}} \put(90.64,
46.96){\circle *{1.2}} \put(94.68, 45.87){\circle *{1.2}}
\put(98.82, 45.22){\circle *{1.2}} \put(103.00, 45.00){\circle
*{1.2}} \put(107.18, 45.22){\circle *{1.2}} \put(111.32,
45.87){\circle *{1.2}} \put(115.36, 46.96){\circle *{1.2}}
\put(119.27, 48.46){\circle *{1.2}} \put(123.00, 50.36){\circle
*{1.2}} \put(126.51, 52.64){\circle *{1.2}} \put(129.77,
55.27){\circle *{1.2}} \put(132.73, 58.23){\circle *{1.2}}
\put(135.36, 61.49){\circle *{1.2}} \put(137.64, 65.00){\circle
*{1.2}} \put(139.54, 68.73){\circle *{1.2}} \put(141.04,
72.64){\circle *{1.2}} \put(142.13, 76.68){\circle *{1.2}}
\put(142.78, 80.82){\circle *{1.2}}
\end{picture}

\vskip -1.7cm
\begin{center}
\begin{minipage}{12cm}\vspace{-1.4cm}
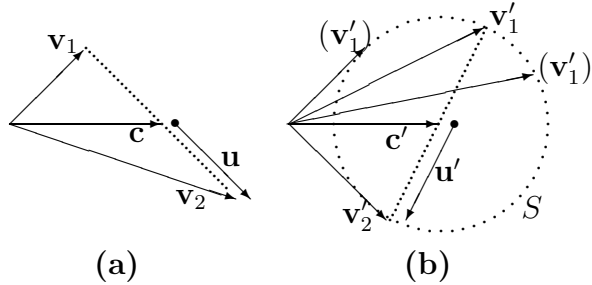
{Figure~2: Constraints imposed by the energy and momentum
conservation laws.  }
\end{minipage}
\end{center}

\vskip -0.5cm Unfortunately, the cross section in (\ref{equality})
and (\ref{sigma1}) is ill-defined. Notice that the energy and
momentum conservation laws state that (assuming every particle to
have the same mass for simplicity)
\begin{equation} \label{conservation}
{\bf c}={\bf c}' \quad{\rm and}\quad |{\bf u}|=|{\bf u}'|\equiv u,
\end{equation}
where $2{\bf c}={\bf v}_1+{\bf v}_2$, $2{\bf c}'={\bf v}'_1+ {\bf
v}'_2$, $2{\bf u} = {\bf v}_2-{\bf v}_1$ and $2{\bf u}'={\bf
v}'_2- {\bf v}'_1$. Fig.~2a shows how ${\bf v}_1$ and ${\bf v}_2$
determine ${\bf c}$ and ${\bf u}$, while Fig.~2b shows how $\bf c$
and $u\equiv |{\bf u}|$ form four constraint constants on ${\bf
v}'_1$ and ${\bf v}'_2$. Notably, ${\bf v}'_1$, as well as ${\bf
v}'_2$, falls on the spherical shell $S$ of radius $u$ in the
velocity space, which will be called the accessible shell. With
these constraints in mind, two problems associated with the
definition (\ref{sigma1}) will surface by themselves. The first is
that, after $d{\bf v}'_1$ is specified, specifying $d{\bf v}'_2$
in the definition is a work overdone (since ${\bf v}'_1$ and ${\bf
v}'_2$ are not independent of each other). The second is that the
cross section should be defined in reference to surface elements
rather than to volume elements.

To see the second problem aforementioned more vividly, let's
consider $d{\bf v}'_1$ shown in Fig.~3a, which is cube-shaped with
equal sides $l$. If we let $\rho$ denote the area density of
particles on the accessible shell caused by unit flux of type-1
particles, the number of type-1 particles found in $d{\bf v}'_1$
can be expressed as $N\approx \rho l^2$. Then, the cross section
defined by (\ref{sigma1}) is equal to, with $d{\bf v}_2^\prime$
omitted,
\begin{equation}\label{v-to-s}
\sigma=\frac N{d{\bf v}_1^\prime}=\frac{\rho l^2} {l^3}=
\frac{\rho}{l},
\end{equation}
which depends on $l$ and tends to infinity as the cube becomes
smaller and smaller. Nevertheless, if $d{\bf v}_1^\prime$ is
chosen to be a slim box in Fig.~3b and the box becomes slimmer and
slimmer, then $\sigma$ tends to zero; if $d{\bf v}_1^\prime$ is
chosen to be a short box in Fig.~3c and the box becomes shorter
and shorter, $\sigma$ tends to infinity again. These
representative examples inform us that the cross section defined
by (\ref{sigma1}), and thus the time reversibility expressed by
(\ref{equality}), does not mean anything.

\hspace{1.7cm} \setlength{\unitlength}{0.012in}
\begin{picture}(300,138)

\multiput(85,87)(90,0){3}{\makebox(8,8)[c]{$S$}}
\put(57,42){\makebox(8,8)[c]{\bf (a)}}
\put(147,42){\makebox(8,8)[c]{\bf (b)}}
\put(237,42){\makebox(8,8)[c]{\bf (c)}}

\put(46,105){\makebox(30,8)[c]{$d{\bf v}_1^\prime$}}
\put(136,118){\makebox(30,8)[c]{$d{\bf v}_1^\prime$}}
\put(226,105){\makebox(30,8)[c]{$d{\bf v}_1^\prime$}}

\put(145,75){\framebox(10,40){}} \put(220,90){\framebox(40,10){}}
\put(55,90){\framebox(10,10){}}
\multiput(95.00,60.00)(90,0){3}{\circle*{1}}
\multiput(95.00,60.00)(90,0){3}{\circle*{1}}
\multiput(94.89,62.82)(90,0){3}{\circle*{1}}
\multiput(94.55,65.61)(90,0){3}{\circle*{1}}
\multiput(93.98,68.38)(90,0){3}{\circle*{1}}
\multiput(93.20,71.08)(90,0){3}{\circle*{1}}
\multiput(92.20,73.72)(90,0){3}{\circle*{1}}
\multiput(90.99,76.27)(90,0){3}{\circle*{1}}
\multiput(89.58,78.71)(90,0){3}{\circle*{1}}
\multiput(87.98,81.03)(90,0){3}{\circle*{1}}
\multiput(86.20,83.21)(90,0){3}{\circle*{1}}
\multiput(84.25,85.24)(90,0){3}{\circle*{1}}
\multiput(82.14,87.11)(90,0){3}{\circle*{1}}
\multiput(79.88,88.80)(90,0){3}{\circle*{1}}
\multiput(77.50,90.31)(90,0){3}{\circle*{1}}
\multiput(75.00,91.62)(90,0){3}{\circle*{1}}
\multiput(72.41,92.73)(90,0){3}{\circle*{1}}
\multiput(69.74,93.62)(90,0){3}{\circle*{1}}
\multiput(67.00,94.29)(90,0){3}{\circle*{1}}
\multiput(64.22,94.74)(90,0){3}{\circle*{1}}
\multiput(61.41,94.97)(90,0){3}{\circle*{1}}
\multiput(58.59,94.97)(90,0){3}{\circle*{1}}
\multiput(55.78,94.74)(90,0){3}{\circle*{1}}
\multiput(53.00,94.29)(90,0){3}{\circle*{1}}
\multiput(50.26,93.62)(90,0){3}{\circle*{1}}
\multiput(47.59,92.73)(90,0){3}{\circle*{1}}
\multiput(45.00,91.62)(90,0){3}{\circle*{1}}
\multiput(42.50,90.31)(90,0){3}{\circle*{1}}
\multiput(40.12,88.80)(90,0){3}{\circle*{1}}
\multiput(37.86,87.11)(90,0){3}{\circle*{1}}
\multiput(35.75,85.24)(90,0){3}{\circle*{1}}
\multiput(33.80,83.21)(90,0){3}{\circle*{1}}
\multiput(32.02,81.03)(90,0){3}{\circle*{1}}
\multiput(30.42,78.71)(90,0){3}{\circle*{1}}
\multiput(29.01,76.27)(90,0){3}{\circle*{1}}
\multiput(27.80,73.72)(90,0){3}{\circle*{1}}
\multiput(26.80,71.08)(90,0){3}{\circle*{1}}
\multiput(26.02,68.38)(90,0){3}{\circle*{1}}
\multiput(25.45,65.61)(90,0){3}{\circle*{1}}
\multiput(25.11,62.82)(90,0){3}{\circle*{1}}
\end{picture}
\vskip-2.0cm
\begin{center}
\begin{minipage}{12cm}\vspace{-2.0cm}
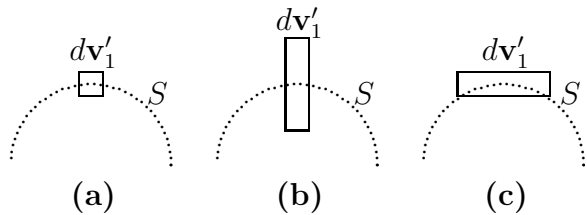
{Figure~3: In defining the cross section, the velocity volume
elements $d{\bf v}_1^\prime$ may take on different shapes.}
\end{minipage}
\end{center}

\vskip-0.7cm As a matter of fact, the above problem can be
examined more briefly. For any definite velocities ${\bf v}_1$ and
${\bf v}_2$, the six components of ${\bf v}'_1$ and ${\bf v}'_2$
are under four constraint equations imposed by the energy-momentum
conservation laws. This literally means that the free space of
${\bf v}'_1$ and ${\bf v}'_2$ is just two-dimensional and there
are no enough degrees of freedom allowing us to define a cross
section in reference to six-dimensional volume elements.

\section{No time-reversibility in terms of velocity volumes}
Another form of time reversibility is simultaneously employed in
such textbooks\cite{reif}:
\begin{equation}\label{timerev2} d{\bf
v}_1 d{\bf v}_2=d{\bf v}'_1 d{\bf v}'_2. \end{equation} It should
be noted that there is a conceptual conflict between
(\ref{equality}) and (\ref{timerev2}). In connection with
(\ref{equality}),  when incident particles have two definite
velocities the velocities of scattered particles are allowed to
distribute over almost the entire velocity space; whereas, in
connection with (\ref{timerev2}), an infinitesimal velocity range
of incident particles strictly corresponds to another
infinitesimal velocity range of scattered particles. Nevertheless,
for purposes of this paper, we shall leave this conflict alone.

The mathematical proof of (\ref{timerev2}) goes as follows. First,
\begin{equation} \label{j1}
d{\bf v}_1 d{\bf v}_2=\|J\|d{\bf c}d{\bf u}\quad {\rm with} \quad
\|J\|=\left\| \frac{\partial ({\bf v}_1,{\bf v}_2)} {\partial
({\bf c},{\bf u})} \right\|. \end{equation} Then,
\begin{equation}\label{j2}
d{\bf v}'_1 d{\bf v}'_2=\|J'\|d{\bf c}'d{\bf u}'\quad {\rm with}
\quad \|J'\|=\left\| \frac{\partial ({\bf v}'_1,{\bf v}'_2)}
{\partial ({\bf c}',{\bf u}')} \right\|. \end{equation} In view of
that
\begin{equation}\label{j3}
\|J\|=\|J'\|,\quad {\bf c}\equiv {{\bf c}'} \quad d{\bf u}=d{\bf
u}',\end{equation} we obtain (\ref{timerev2}).

Unfortunately again, the formulation given above also involves
errors. All equations in (\ref{j1}), (\ref{j2}) and (\ref{j3})
hold except $d{\bf u}=d{\bf u}'$. Referring to Fig.~2, we find
that when ${\bf u}$ is a definite vector, ${\bf u}'$ distributes
over a spherical shell, pointing in any direction. This means that
if $d{\bf u}=u^2dud\Omega_{\bf u}$ is an infinitely thin
line-shaped volume element (say, $d\Omega_{\bf u}$ is
infinitesimal while $du$ finite), the corresponding volume element
$d{\bf u}'$ will be a spherical shell with finite thickness. It is
then obvious that the two elements are not equal in volume.

The issue can be analyzed more economically in terms of variable
transformation. If we identify $v_{1x},v_{1y},v_{1z}$ and
$v_{2x},v_{2y},v_{2z}$ as six variables and identify
$v'_{1x},v'_{1y},v'_{1z}$ and $v'_{2x},v'_{2y},v'_{2z}$ as six new
variables, then we have \begin{equation} \label{jacobian} d{\bf
v}_1 d{\bf v}_2=\|\hat J\| d{\bf v}'_1 d{\bf v}'_2. \end{equation}
What has been proven by equations (\ref{j1}), (\ref{j2}) and
(\ref{j3}) is nothing but $\|\hat J\|=1$. However, in order for
(\ref{jacobian}) to make sense, there must exist six independent
equations connecting those variables. In our case, we have four
equations only. That is to say, the `variable transformation' is
incomplete and expression (\ref{jacobian}) is not truly
legitimate.

It is instructive to look at the issues discussed in the last and
this sections in a unifying way. If there were no single
constraint equation, defining the cross section could make sense;
if there were six independent constraint equations, defining the
Jacobian would be meaningful. Since there are four and only four
constraint equations, neither the cross section nor the Jacobian
can be defined.

\section{Formulation of beam-to-beam collisions}
It is now rather clear that the concepts and methodologies of
Boltzmann's theory are in need of reconsideration.

For instance, one of the major steps in deriving Boltzmann's
equation is to identify $f({\bf v}'_1)d{\bf v}'_1$ and $f({\bf
v}'_2)d{\bf v}'_2$ as two definite beams and then to determine how
many beam-1 particles will emerge between ${\bf v}_1$ and ${\bf
v}_1+d{\bf v}_1$ due to collisions of the two beams. As has been
shown, this context leads to nothing but an absurd result: the
number of such emerging particles actually depends on the size and
shape of $d{\bf v}_1$, varying drastically from zero to infinity.

In what follows, we shall propose a new context to do the job.
Surprisingly, the formulation will reveal some of deep-rooted
properties of statistical mechanics.

To involve less details, we adopt the following assumptions: (i)
The zeroth-order, collisionless, distribution function of the gas
is completely known. (ii) Each particles, though belonging to the
same species, is still distinguishable (which is possible in terms
of classical mechanics). (iii) No particle collides twice or more.
(General treatments can be accomplished along this
line\cite{chen99}.)

Referring to Fig.~4, we consider that two typical beams, denoted
by $ f^{(0)}_1({\bf v}'_1) d{\bf v}'_1$ and $f^{(0)}_2({\bf v}'_2)
d{\bf v}'_2$, collide with each other, and suppose that a particle
detector has been placed somewhere in the region. Let $\Delta S$
be the entry area of the detector and $\Delta N_1$ be the number
of the beam-1 particles entering the detector within the velocity
range $\Delta v_1 v_1^2 \Delta \Omega_1$ during $d t$. Since any
beam of the system can be regarded as the first beam, or the
second beam, aforementioned, the total distribution function due
to collisions is, at the detector entry,
\begin{equation}\label{deltaN} f^{(1)}_1(t,{\bf r}, {\bf v}_1,\Delta
S,\Delta v_1,\Delta \Omega_1 )\approx \frac {\sum_1\sum_2\Delta
N_1}{(\Delta S v_1 d t)(\Delta v_1 v_1^2 \Delta
\Omega_1)},\end{equation} in which ${\bf r}$ is the representative
position of $\Delta S$ and ${\bf v}_1$ is the representative
velocity of $\Delta v_1 v_1^2 \Delta \Omega_1$. According to the
customary thought, when $\Delta S$, $\Delta v_1$ and $\Delta
\Omega_1$ shrink to zero simultaneously, expression (\ref{deltaN})
stands for the `true distribution function' there; for reasons to
be clear a bit later, we shall, in the following formulation,
assume that $\Delta S$ and $\Delta v_1$ are infinitely small while
$\Delta \Omega_1$ is kept fixed and finite (though rather small).
The spatial region $-\Delta\Omega_1$, that has been shaded in
Fig.~4 and `opposite' to the velocity solid-angle range $\Delta
\Omega_1$ in (\ref{deltaN}), will be called the effective cone. It
is intuitively obvious that the particles that collide somewhere
in the effective cone and move, after the collision, toward the
detector entry along their free trajectories will contribute to
$\Delta N_1$. (Even if such particles are allowed to collide
again, some of them will still arrive at the detector freely,
which means the concept of effective cone holds its significance
rather generally.)

\hspace{0.7cm} \setlength{\unitlength}{0.020in}
\begin{picture}(200,98)
\multiput(85,45)(0,4){2}{\vector(3,1){15}}
\multiput(85,70)(0,4){3}{\vector(3,-1){15}}

\multiput(110,40)(12,0){2}{\line(1,0){8}}
\multiput(110,40)(20,0){2}{\line(0,-1){12}}

\multiput(120,40)(-0.3,1.2){34}{\circle*{0.7}}
\multiput(120,40)(0.3,1.2){34}{\circle*{0.7}}

\multiput(119,48)(1.5,0.4){3}{\circle*{0.7}}
\multiput(118,52)(1.5,0.4){4}{\circle*{0.7}}
\multiput(117,56)(1.5,0.4){5}{\circle*{0.7}}
\multiput(116,60)(1.5,0.4){7}{\circle*{0.7}}
\multiput(115,64)(1.5,0.4){9}{\circle*{0.7}}
\multiput(114,68)(1.5,0.4){10}{\circle*{0.7}}
\multiput(113,72)(1.5,0.4){11}{\circle*{0.7}}

\put(132,30){\makebox(20,8)[l]{Detector}}
\put(110,80){\makebox(20,8)[c]{$\small -\Delta\Omega_1$}}
\put(63,42){\makebox(20,8)[c]{$f^{(0)}_1d{\bf v}'_1 $}}
\put(63,75){\makebox(20,8)[c]{$f^{(0)}_2d{\bf v}'_2 $}}
\end{picture}

\vspace{-29pt}\begin{center}
\begin{minipage}{12cm} {
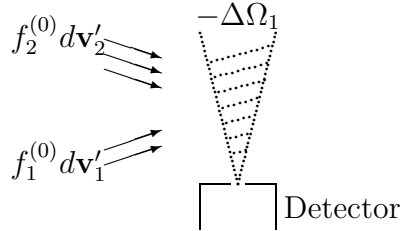
\vskip-0.3cm Figure~4: Two beams collide and a particle detector
is placed in the region.} \end{minipage}
\end{center}

Observing the colliding beams in the center-of-mass reference
frame, we find that the number of collisions in a volume element
$d{\bf r}'$, which is located inside the effective cone $-\Delta
\Omega$, can be represented by, as in Boltzmann's theory,
 \begin{equation}
\label{twobeams} [d{\bf r}'f^{(0)}_1({\bf v}'_1)d{\bf v}'_1]
[f^{(0)}_2({\bf v}'_2)d{\bf v}'_2] [2u\sigma_c({\bf u}',{\bf u}) d
\Omega_c d t ],\end{equation} where $\Omega_c$ is the solid angle
between ${\bf u}'$ and ${\bf u}$, and $\sigma_c({\bf u}',{\bf u})$
is the cross section in the center-of-mass frame. By integrating
(\ref{twobeams}) over the effective cone and taking account of all
the particles that are registered by the detector, the right side
of (\ref{deltaN}) becomes
\begin{equation}\label{entry} \int_{-\Delta \Omega_1}d{\bf r}'
\int_{\Delta
v_1\Delta \Omega_1} d\Omega_c \int d{\bf v}'_1\int d{\bf v}'_2
\frac{2u\sigma_c({\bf u}',{\bf u}) f^{(0)}_1({\bf v}'_1)
f^{(0)}_2({\bf v}'_2)} {(|{\bf r}-{\bf r}'|^2 \Delta \Omega_0
v_1)\dot (v_1^2\Delta v_1 \Delta \Omega_1 )},
\end{equation} where $\Delta \Omega_0$ is the solid-angle range
formed by a representative point in $d{\bf r}'$ (as the apex) and
the detector entry area $\Delta S$ (as the base). In view of that
$\Delta S$ is truly small and $\Delta\Omega_1$ is fixed and finite
by our assumption, we know that $\Delta\Omega_0 \ll
\Delta\Omega_1$ and every particle starting its free journey from
the effective cone and entering the detector can be treated as one
emerging within $\Delta \Omega_1$. Then, we can rewrite expression
(\ref{entry}) as, with help of the variable transformation from
$({\bf v}'_1,{\bf v}'_2)$ to $({\bf c}',{\bf u}')$,
\begin{equation}\label{entry1} \int_{-\Delta \Omega_1}d{\bf
r}'\int_{\Delta v_1\Delta \Omega_0} d\Omega_c \int d{\bf c}' \int
d{\bf u}'\|\tilde J\| \frac{2u\sigma_c({\bf u}',{\bf u})
f^{(0)}_1({\bf c}'-{\bf u}') f^{(0)}_2({\bf c}'+{\bf u}')} {(|{\bf
r}-{\bf r}'|^2 \Delta \Omega_0 v_1)\dot (v_1^2\Delta v_1 \Delta
\Omega_1 )},
\end{equation} in which $\|\tilde J\|\equiv \partial({\bf v}'_1,
{\bf v}'_2) / \partial({\bf c}',{\bf u}' )$ and the subindex
$\Delta v_1\Delta \Omega_1$ has been replaced by $\Delta v_1\Delta
\Omega_0$. In view of the energy-momentum conservation laws, we
arrive at
\begin{equation}\label{entry2}
\begin{array}{l}\displaystyle \int_{-\Delta \Omega_1}d{\bf
r}'  \int d{\bf c}' \int_{4\pi} d\Omega_{{\bf u}'}\int_{\Delta
v_1\Delta \Omega_0}d\Omega_c\int du  \vspace{3pt}\\
\displaystyle \hspace{4cm}\cdot \|\tilde J\| \frac{2u\sigma_c({\bf
u}',{\bf u}) f^{(0)}_1({\bf c}'-{\bf u}') f^{(0)}_2({\bf c}'+{\bf
u}')} {(|{\bf r}-{\bf r}'|^2 \Delta \Omega_0 v_1)\dot (v_1^2\Delta
v_1 \Delta \Omega_1 )}, \end{array} \end{equation}

\hspace{1.3cm} \setlength{\unitlength}{0.013in}
\begin{picture}(200,117)
\put(155,93){\makebox(35,8)[l]{${\bf c}$}}
\put(165,45){\makebox(35,8)[l]{$\bf u$}}
\put(220,90){\vector(-1,-1){60}}
\multiput(220,90)(-1.9,-2.3){29}{\circle*{1.2}}
\put(70,90){\line(1,0){150}} \put(217.5,90.5){\vector(1,0){1}}
\multiput(70,90)(3.0,-2.3){34}{\circle*{1.2}}
\multiput(70,90)(3.0,-1.7){37}{\circle*{1.2}}
\multiput(170,14.7)(1.6,2.4){6}{\circle*{1.2}}
\multiput(154,42.4)(-1.6,-2.4){5}{\circle*{1.2}}
\put(72,63){\makebox(35,8)[l]{$\Delta\Omega_0$}}
\put(207,65){\makebox(35,8)[l]{$d\Omega_c$}}
\put(140,14){\makebox(35,8)[l]{$\Delta v_1$}}

\put(143.22,53.87){\circle*{1}} \put(144.80,50.69){\circle*{1}}
\put(146.52,47.57){\circle*{1}} \put(148.36,44.53){\circle*{1}}
\put(150.32,41.57){\circle*{1}} \put(152.41,38.70){\circle*{1}}
\put(154.62,35.91){\circle*{1}} \put(156.94,33.22){\circle*{1}}
\put(159.38,30.63){\circle*{1}} \put(161.91,28.15){\circle*{1}}
\put(164.56,25.77){\circle*{1}} \put(167.29,23.50){\circle*{1}}
\put(170.13,21.35){\circle*{1}} \put(173.04,19.32){\circle*{1}}
\put(176.04,17.42){\circle*{1}} \put(179.12,15.64){\circle*{1}}
\put(182.27,14.00){\circle*{1}} \put(185.49,12.48){\circle*{1}}

\end{picture}

\begin{center}
\begin{minipage}{12cm} {
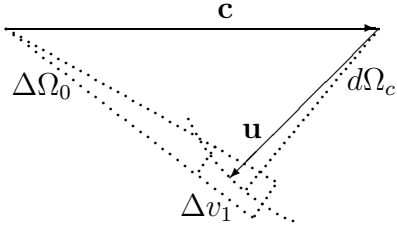
\vskip-0.54cm Figure~5: The relation between the velocity element
$v_1^2\Delta v_1 \Delta\Omega_0 $ and the velocity element $u^2 du
d\Omega_c $.} \end{minipage}
\end{center}
\vspace{0pt}

By examining the situation shown in Fig.~5, which is drawn in the
velocity space, the following relation can be found out:
\begin{equation} \int_{\Delta v_1 \Delta \Omega_0}u^2 du d\Omega_c
\cdots \approx v_1^2\Delta v_1 \Delta\Omega_0\cdots
.\end{equation} Therefore, the distribution function due to
collisions is equal to \begin{equation}\label{final}
\begin{array}{l}\displaystyle f^{(1)}_1(t,{\bf
r}, v_1, \Delta \Omega_1)=\frac{1}{v_1 \Delta \Omega_1}
\displaystyle \int_{-\Delta \Omega_1}d{\bf r}' \int d{\bf c}'
\int_{4\pi} d\Omega_{{\bf u}'}\vspace{3pt}\\\hspace{4.5cm}
\displaystyle \cdot\frac{2\|\tilde J\|\sigma_c({\bf u}',{\bf u})
f^{(0)}_1({\bf c}'-{\bf u}') f^{(0)}_2({\bf c}'+{\bf u}')} {u|{\bf
r}-{\bf r}'|^2},
\end{array} \end{equation}
in which $\bf u$ is determined by ${\bf u}={\bf c}-{\bf v}_1$
(${\bf v}_1$ is in principle along the direction of ${\bf r}-{\bf
r}'$) and ${\bf u}'$ by $u$ and $\Omega_{{\bf u}'}$. It should be
noted that $f^{(1)}_1$ in (\ref{final}) differs from that in
(\ref{deltaN}) in the sense that $\Delta S$ and $\Delta v_1$ cease
to be arguments and $v_1$ takes the place of ${\bf v}_1$.

If the zeroth-order distribution functions depend on time and
space, the replacement
\begin{equation} f^{(0)}_1({\bf c}'-{\bf u}')f^{(0)}_2({\bf c}'+
 {\bf u}')\rightarrow
f^{(0)}_1(t',{\bf r}',{\bf c}'-{\bf u}')f^{(0)}_2(t',{\bf r}',{\bf
c}'+{\bf u}')\end{equation} needs to be taken, where $t'=t-|{\bf
r}-{\bf r}'|/v_1$ stands for the time delay.

Since $\Delta\Omega_1$ has been set finite, expression
(\ref{final}) is nothing but the distribution function averaged
over $\Delta\Omega_1$, which is not good enough according to the
standard theory. At first glance, if we let $\Delta\Omega_1$
become smaller and smaller, expression (\ref{final}) will finally
represent the true distribution function there. However, the
discussion after (\ref{entry}) has shown that if $\Delta \Omega_1$
shrinks to zero the definition of the effective cone will collapse
and thus the entire formalism will no longer be valid.

A careful inspection tells us that, if we assumed $\Delta S$ to be
finite and $\Delta \Omega_1$ to be infinitesimal at the starting
point, a different average distribution function, averaged over
$\Delta S$, would be obtained. The fact that no distribution
function can be determined if $\Delta \Omega_1$ and $\Delta S$
simultaneously take on their infinitesimal values suggests that
even in classical statistical mechanics there also exists an
uncertainty principle. The connection between this uncertainty
principle and the uncertainty principle in quantum mechanics
remains to be seen.

In view of that the integration in (\ref{final}) is carried out
over an effective region defined by free trajectories of
particles, this methodology has been named as a path-integral
approach\cite{chen99}. As (\ref{final}) has partially shown,
anything taking place in the effective region can make direct
impact along `free' paths and any macroscopic structures,
continuous or not, will help shape microscopic structures
elsewhere along `free' paths. Concepts like the aforementioned,
though appearing to be foreign from the viewpoint of differential
approach, are in harmony with what takes place in realistic gases.

\section{Summary}

It has been shown that, when we concern ourselves with
beam-to-beam collisions, part of the system's information has,
knowingly or not, been disregarded, and the information loss is
characterized by the fact that we have four and only four
constraint equations (rather than six). Since statistical
mechanics deals with beam-to-beam collisions, the
time-reversibility related to each individual collision becomes
irrelevant from the very beginning and will not reemerge at any
later stage.

By proposing a new context in which what can be really measured in
an experiment is of central interest, beam-to-beam collisions have
been reformulated. The new formulation reveals that only
distribution functions averaged over certain finite ranges make
good physical sense.

Apparently, this paper raises many difficult questions related to
the very foundation of statistical physics. Reference papers can
be found in the regular and e-print
literature\cite{chen02,chen99,chen05}.

\noindent{\bf Acknowledgments:} The author is very grateful to
professor Oliver Penrose, who reminded me of possibility of
defining the cross section of delta-function type. Such definition
and its possible consequences are discussed in Appendix A. This
paper is supported by School of Science, BUAA, PRC.

\appendix

\section*{Appendix A. An alternative cross section}

As revealed in Sect.~3, the original cross section $\sigma({\bf
v}_1,{\bf v}_2 \to {\bf v}'_1,{\bf v}'_2) $ defined by
(\ref{sigma1}) cannot be interpreted as a function of the usual
kind due to the existence of the energy-momentum conservation
laws. However, one may still wish to define cross sections in
which the time reversibility related to a collision of two
particles plays a certain role. Interestingly, this can be done
and doing so will help us to find out what kind of problems
Boltzmann's equation really has.

If we adopt the definition of (\ref{sigma1}) and, at the same
time, take the energy-momentum conservation laws into account, we
are led to a cross section of delta-function type
\begin{equation}\label{app1}
\begin{array}{l}\sigma({\bf v}_1,{\bf v}_2 \to {\bf v}_1',{\bf
v}_2') =\displaystyle\eta({\bf v}_1,{\bf v}_2 \to {\bf v}_1',{\bf
v}_2')\cdot \vspace{3pt}\\\hspace{20pt} \delta\left(
\sqrt{(v'_1)^2  +(v'_2)^2}- \sqrt{(v_1)^2 + (v_2)^2}\right)\cdot
\delta^3 \left[({\bf v}'_1 + {\bf v}'_2)-({\bf v}_1 + {\bf v}_2)
\right]. \end{array} \end{equation} To work out the physical
meaning of $\eta$, we integrate the above expression over a finite
but small volume element $\Delta {\bf v}_1'\Delta {\bf v}_2'$
\begin{equation}\label{app2}\begin{array}{l}\displaystyle
\int_{\Delta {\bf v}_1'\Delta {\bf v}_2'}\eta({\bf v}_1,{\bf v}_2
\to {\bf v}_1',{\bf v}_2')\cdot\vspace{3pt}\\ \delta\left(
\sqrt{(v'_1)^2 +(v'_2)^2}- \sqrt{(v_1)^2 + (v_2)^2}\right)\cdot
\delta^3 \left[({\bf v}'_1 + {\bf v}'_2)-({\bf v}_1 + {\bf v}_2)
\right] d{\bf v}_1'd{\bf v}_2'.
\end{array}\end{equation}
As shown in Sect.~3, these delta-functions define an accessible
shell, denoted by $S$ there, and the integration becomes
\begin{equation}\label{eta-s}
\eta({\bf v}_1,{\bf v}_2 \to {\bf v}_1',{\bf v}_2') \Delta S,
\end{equation}
where $\Delta S$ is enclosed by $\Delta {\bf v}_1'\Delta {\bf
v}_2'$ (actually by one of $\Delta {\bf v}_1'$ and $\Delta {\bf
v}_2'$ since the two are not independent of each other). By
comparing this with the cross section defined in the
center-of-mass system $\sigma_c({\bf u},{\bf u}') d\Omega $, it is
clear that
\begin{equation}\label{eta}
\eta({\bf v}_1,{\bf v}_2 \to {\bf v}_1',{\bf v}_2') =\sigma_c({\bf
u},{\bf u}')/u^2 .\end{equation} The above formalism has
illustrated that the introduced delta-functions make good sense as
long as they are integrated over an adequate volume (finite or
infinitely large).
 Similarly,
\begin{equation} \label{app11}
\begin{array}{l}\sigma({\bf v}_1',{\bf v}_2' \to {\bf v}_1,{\bf
v}_2) =\displaystyle \eta({\bf v}_1',{\bf v}_2' \to {\bf v}_1,{\bf
v}_2)\cdot  \vspace{3pt}\\ \hspace{20pt}\delta\left( \sqrt{(v_1)^2
+(v_2)^2}- \sqrt{(v_1')^2 + (v_2')^2}\right)\cdot \delta^3
\left[({\bf v}_1 + {\bf v}_2)-({\bf v}_1' +{\bf v}_2')
\right].\end{array}\end{equation} With help of (\ref{eta}), we see
that
\begin{equation}
\eta({\bf v}_1,{\bf v}_2 \to {\bf v}'_1,{\bf v}'_2) = \eta({\bf
v}'_1,{\bf v}'_2 \to {\bf v}_1,{\bf v}_2).
\end{equation}
This reflects the fact that all collisions, including the original
collision and the inverse collision, are of head-on type in the
center-of-mass system.

It is now in order to find out whether or not the newly defined
cross section is relevant to Boltzmann's equation. We first
examine the particles leaving $d{\bf r}d{\bf v}_1$ during $dt$
because of collisions. Following the textbook treatment, the
number of collisions is represented by
\begin{equation} [2uf({\bf v}_1)d{\bf v}_1][f({\bf v}_2)d{\bf
r}d{\bf v}_2] \sigma({\bf v}_1,{\bf v}_2 \to {\bf v}'_1,{\bf
v}'_2)dt.
\end{equation}
Integrating it over ${\bf v}_1'$, ${\bf v}_2'$ and ${\bf v}_2$
yields, by virtue of (\ref{app1}), (\ref{eta-s}) and (\ref{eta}),
\begin{equation} dtd{\bf r}d{\bf v}_1\int  2uf({\bf
v}_1)f({\bf v}_2)\sigma_c({\bf u},{\bf u}')d{\bf v}_2 d\Omega.
\end{equation}
Dividing it by $dtd{\bf r}d{\bf v}_1$, we obtain the collision
number per unit time and unit phase volume
\begin{equation}
\int  2uf({\bf v}_1)f({\bf v}_2)\sigma_c({\bf u},{\bf u}')d{\bf
v}_2 d\Omega.
\end{equation}
If we adopt the assumption that the above number is identical to
the number of particles leaving the unit phase volume per unit
time because of collisions (though a different conclusion is
offered in Ref.~2), we find that the above derivation is entirely
consistent with that in the standard approach.

Then, we examine the particles entering $d{\bf r}\Delta {\bf v}_1$
during $dt$ because of collisions ($\Delta{\bf v}_1$ has been set
finite for a reason that will be clarified). To make our
discussion a bit simpler, it is assumed that there are only two
incident beams
\begin{equation}\label{twobeams}
f({\bf v}_1') \Delta{\bf v}_1' \quad {\rm and}\quad f({\bf v}_2')
\Delta{\bf v}_2'.\end{equation} Again, following the standard
approach, we know that the collision number caused by the two
beams within $d{\bf r}$ during $dt$ is
\begin{equation}
 [2uf({\bf v}_1')\Delta{\bf v}_1']\cdot[f({\bf v}_2')d{\bf
r}\Delta{\bf v}_2']\cdot \sigma({\bf v}_1',{\bf v}_2' \to {\bf
v}_1,{\bf v}_2)dt.
\end{equation}
At this point, a sharp question arises. Can we identify these
particles as those entering $d{\bf r}\Delta {\bf v}_1$ during
$dt$? The answer is apparently a negative one. As one thing, only
a small fraction of the scattering particles will enter
$\Delta{\bf v}_1$, and the following integration needs to be done:
\begin{equation}\label{enter1}
\int_{\Delta {\bf v}_1\Delta {\bf v}_2} 2uf({\bf v}_1')\Delta{\bf
v}_1'\cdot f({\bf v}_2')d{\bf r}\Delta{\bf v}_2'\cdot \sigma({\bf
v}_1',{\bf v}_2' \to {\bf v}_1,{\bf v}_2) dtd{\bf v}_1d{\bf v}_2,
\end{equation}
in which $\Delta {\bf v}_1$ has to be finite, since the integrand
contains delta-functions, while $\Delta {\bf v}_2$ can be
infinitely large. If we are interested in knowing the number of
the entering particles per unit phase volume and unit time, we are
supposed to evaluate
\begin{equation}\label{limit}
\lim\limits_{dtd{\bf r}\Delta{\bf v}_1\to 0} \frac{\Delta
N}{dtd{\bf r}\Delta{\bf v}_1},
\end{equation}
where $\Delta N$ is nothing but expression (\ref{enter1}).
Unfortunately, there are problems. As one thing, it has just been
pointed out that $\Delta{\bf v}_1$ has to be finite. As another,
expression (\ref{v-to-s}) and the accompanying discussion have
shown that the limit expressed by (\ref{limit}) does not exist.

The above discussion, verifiable with numerical computation,
suggests that (i) The methodology of determining particles
entering a phase volume element must be very different from that
of determining particles leaving a phase volume element, namely
there is no symmetry. (ii) Taking limits $dt\to 0$, $d{\bf r}\to
0$ and $d{\bf v}_1\to 0$, simultaneously or not, can create
unexpected problems. (iii) In order to formulate the collisional
dynamics, new concepts and new methodologies are in need.


\begin{thebibliography}{99}

\bibitem{reif} F.~Reif, {\it Fundamentals of
Statistical and Thermal Physics}, (McGraw-Hill Book Company, 1965,
1987, 1988 in English and German); L.~E.~Reichl {\it A Modern
Course in Statistical Physics}, 2nd ed., (John Wiley and Sons, New
York, 1998).
\bibitem{chen02} C.~Y.~Chen, Il Nuovo Cimento B {\bf V117B}, 177
(2002), which reveals some problems of Boltzmann's equation;
C.~Y.~Chen, Journal of Physics A, 6589, (2002), which discusses
something that can actually challenge Liouville's theorem in
quantum statistical physics.
\bibitem{chen99} C.~Y.~Chen, {\it Perturbation Methods and Statistical
Theories}, in English, (International Academic Publishers,
Beijing, 1999).
\bibitem{chen05} C.~Y.~Chen, cond-mat/0412396;
physics/0312043, 0311120, 0305006, 0010015, 0006033, 0006009,
9908062; quant-ph/0009023, 0009015, 9911064, 9907058.

\end{thebibliography}
\end{document}